\documentstyle[12pt,aasms4]{article}
\def\simgt{\lower.5ex\hbox{$\; \buildrel > \over \sim \;$}}
\def\simlt{\lower.5ex\hbox{$\; \buildrel < \over \sim \;$}}

\begin{document}
\title{Convective-Dynamical Instability in Radiation-Supported Accretion Disks}

\author{Paola Pietrini}
\affil{ Dipartimento di Astronomia e Scienza dello Spazio, Universit\'a 
        di Firenze, Largo E. Fermi 5, 50125 Firenze, Italy}    
\and
\author{Julian H. Krolik}

\affil{Department of Physics and Astronomy, Johns Hopkins University, 
	   Baltimore, MD 21218}

\begin{abstract}

    We study radiation-hydrodynamical normal modes of radiation-supported
accretion disks in the WKB limit.  It has long been known that in the
large optical depth limit the standard equilibrium is unstable to
convection.  We study how the growth rate depends on location within
the disk, optical depth, disk rotation, and the way in which the local
dissipation rate depends on density and pressure.  The greatest growth
rates are found near the disk surface.  Rotation stabilizes vertical
wavevectors, so that growing modes tend to have nearly-horizontal
wavevectors.  
Over the likely range of optical depths, the linear growth 
rate for convective instability has only a weak dependence on disk opacity. 
Perturbations to the dissipation have little effect on convective mode growth 
rates, but can cause growth of radiation sound waves.

\end{abstract}

\section{Introduction}

     Shakura \& Sunyaev (1973) predicted that the inner portions of accretion
disks that extend into relativistically-deep gravitational potentials should
be radiation pressure-dominated when the accretion rate is greater than
a modest fraction of the Eddington rate.  In that regime, they found that
disks could achieve hydrostatic balance in the vertical direction if the
local dissipation rate were proportional to the local mass density.  Given
that assumption, upward radiation flux could support the disk matter against
gravity if the density were essentially constant as a function of height
(falling sharply to zero at the top surface) and the radiation pressure
fell gradually from the disk midplane to the surface.

     Soon after this equilibrium was discovered, it was found to suffer from
several sorts of instabilities.  Lightman \& Eardley (1974) pointed out
that if the
viscous stress is proportional to the total pressure (in this case, dominated
by radiation), perturbations with radial wavelengths long compared to
the vertical thickness $h$, but short compared to a radius $r$, grow on the
(comparatively long) viscous inflow timescale.  Shakura \& Sunyaev (1976) then
observed that in these conditions perturbations in the same range
of wavelengths would also grow on the (shorter) thermal timescale.
Bisnovatyi-Kogan \& Blinnikov (1977) noticed that if the radiation is
locked to the gas even on short lengthscales (i.e., if, for the purpose
of dynamics, the optical depth is treated as effectively infinite),
such disks should be convectively unstable, for the specific entropy
decreases outward; the linear growth rate for convective ``bubbles" was
worked out by Lominadze \& Chagelishvili (1984).  More recently,
Gammie (1998) has demonstrated that a magnetic field in radiation-supported
disks can catalyze a short-wavelength ($kh \gg 1$) overstable wave mode.  In
view of these instabilities, it has long been a puzzle just what sort of
equilibrium would actually be found in Nature when the accretion rate is high
enough that radiation pressure-domination might be expected (see, e.g. Shapiro,
Lightman \& Eardley 1976; Liang 1977; Coroniti 1981; Svensson \& Zdziarski 1994;
Szuszkiewicz \& Miller 1997; Krolik 1998).

    In this paper, we take a closer look at the nature of the short
wavelength modes in radiation-supported disks without magnetic fields.
Our goal (motivated by a companion work on radiation-hydrodynamics simulations
of such disks: Agol \& Krolik 2000b) is to examine more closely which
modes can be expected to
grow most quickly, what happens when finite optical depth permits some
photon diffusion, and what role, if any, is played by associated
perturbations in the local dissipation rate.

\section{Problem Definition}

     We begin by writing down the equations of non-relativistic
radiation hydrodynamics so
that we may first describe the equilibrium in this language, and
then discuss linear perturbations to this equilibrium.  Because we are
interested in accretion disks, it is convenient to write them in a rotating
frame.  The first is the usual equation of mass
conservation:
\begin{equation}
{\partial \rho \over \partial t} + \nabla \cdot \left(\rho \vec v\right) = 0.
\end{equation}
Our notation is the usual one, in which $\rho$ is the mass density and
$\vec v$ is the fluid velocity.  Next is the the fluid force equation:
\begin{equation}
\rho {\partial \vec v \over \partial t} + \rho \vec v \cdot \nabla \vec v =
   -\nabla p_g + \rho \vec g + (\kappa\rho/c)\vec{\cal F} 
+ 2\rho \vec v \times \vec \Omega - 
\rho v_r (\partial\Omega/\partial\ln r) \hat\phi
\end{equation}
where $p_g$ is the gas pressure, $\vec g$ is the local gravity, $\kappa$
is the opacity per unit mass, $\vec{\cal F}$ is the radiation flux, and
$\Omega$ is the rotation rate of the fluid.  Although all the fluid quantities
are defined in the rotating frame, the radiation quantities
(e.g., $\vec{\cal F}$) are defined in the frame of the local fluid motion,
i.e. including any departures from corotation
(Mihalas \& Mihalas 1984).  Note that we have omitted
magnetic forces.  

    For the two equations describing radiation energy density and momentum
density, we follow Buchler (1979), but write the Lagrangian time derivatives
explicitly, {\it i.e.}, $D/Dt = (\partial /\partial t +\vec v\cdot\nabla)$.
The evolution of radiation energy density $E$ is described by:
\begin{equation}
{\partial E \over \partial t} + \vec v \cdot \nabla E + \nabla\cdot\vec{\cal F}
+ {\bf p_r} : \nabla \vec v + 
{E \over c^2} \nabla \cdot \vec v +{2\over c^2}\left({\partial \vec v \over \partial t}
+ \vec v \cdot \nabla \vec v \right) \cdot \vec{\cal F} = Q.
\end{equation}
Here ${\bf p_r}$ is the radiation pressure tensor and $Q$ is the net local
emissivity.  
Finally, there is the equation describing the time-dependence of the
radiation momentum density $(1/c^2){\cal F}$:
\begin{equation}
{1 \over c}{\partial \vec{\cal F} \over \partial t} + {\vec v \over c} \cdot
\nabla \vec{\cal F} + c \nabla \cdot {\bf p_r} + {1 \over c}\left(\vec{\cal F} \cdot
\nabla \vec v + \vec{\cal F} \nabla \cdot \vec v\right) +
{1 \over c}\left(E {\bf I} + {\bf p_r}\right) \cdot \left( {\partial \vec v
\over \partial t} + \vec v \cdot \nabla \vec v \right) = \vec q .
\end{equation}
In this equation, ${\bf I}$ is the identity matrix and $\vec q$ is the net
rate per unit volume at which photon momentum is created by radiation
(usually negative in the fluid frame because photon momentum is lost due
to opacity, while newly-created photons are usually isotropic in
the fluid frame).

   In the equilibrium, $\partial /\partial t = \vec v = 0$.  To isolate
the effect of radiation support, we also take the extreme limit of
$p_g \ll p_r$.  If we regard the rotation of the disk matter as cancelling
the radial component of gravity, the only non-trivial remark to make about the
equilibrium is that ${\cal F}_z = cg/\kappa$, where $ g=g_z(z,r)$ is the local
vertical component of gravitational acceleration [in a thin disk,
$g_z(z,r)\simeq G M z/r^3$ for central mass $M$].

   Now consider perturbed versions of equations (1) through (4).  In order
to write these perturbations in Fourier-transform form (i.e., for
any quantity $X$, the perturbation is
$\delta{\tilde X} = \delta X \exp\{{i(k_rr+k_z z)-i\omega t}\}$), we
will suppose that the wavevectors obey three conditions:
that $k = (k_z^2 + k_r^2)^{1/2} \ll \kappa \rho$;
that $k_z \gg 1/h$;
and that $k_r \gg 1/r$.
The first limit means that the diffusion approximation applies, i.e.
${\bf p_r} = p_r{\bf I}$, so that $E = 3p_r$.  The second is the WKB
approximation, as applied to variations in both the radial and vertical
directions.  Note that we further restrict our attention to axisymmetric
perturbations.  The condition $k_z h \gg 1$ also means that we can ignore any gradients in the gravity or equilibrium radiation flux.  In addition,
assuming $k_r\gg 1/r$ allows us to neglect the terms in vector divergences
arising from cylindrical geometry.  For example, after Fourier-transforming, 
$\nabla\cdot\delta\vec v$ becomes $ik_r\delta v_r +\delta v_r/r + 
ik_z\delta v_z \simeq ik_r\delta v_r + ik_z\delta v_z$.
We then  find:
\begin{eqnarray}
-i\omega \delta\rho & + & \rho i\vec k \cdot \delta \vec v = 0 \\
-i\omega \rho \delta v_r & = & (\kappa \rho /c) \delta {\cal F}_r +
         2\rho\Omega\delta v_\phi \\
-i\omega\rho \delta v_z & = & (\kappa \rho/c) \delta {\cal F}_z \\
-i\omega\rho \delta v_\phi & = & (\kappa\rho/c)\delta {\cal F}_\phi - 
         (1/2)\rho\Omega \delta v_r \\
-3i\omega \delta p_r  -  3 \rho g \delta v_z & + & ik_z \delta {\cal F}_z + 
ik_r\delta{\cal F}_r  + 4p_r\left(i k_z \delta v_z + i k_r \delta v_r\right) 
-i{2\omega\over c^2}\delta v_z F_z  =
 \delta Q \\
-i(\omega/c)\delta {\cal F}_r & + & ik_r c \delta p_r  +  i k_z (g/\kappa) 
\delta v_r
- 4i(\omega/c)p_r \delta v_r  =  -\kappa\rho \delta {\cal F}_r \\
-i\left({\omega\over c}\right)\delta {\cal F}_z  + ik_z c \delta p_r  &
+ & {g\over\kappa}(2ik_z \delta v_z + ik_r \delta v_r) - 4i{\omega\over c}p_r \delta v_z =
     -cg\delta\rho - \kappa\rho \delta{\cal F}_z
\\
-i(\omega/c)\delta {\cal F}_\phi &+ & i(k_z/c){\cal F}_z\delta v_\phi  -  
4i(\omega/c)p_r\delta v_\phi = -\kappa \rho \delta {\cal F}_\phi 
\end{eqnarray}
The second and third equations in this set are the two components of the
fluid force equation, with the vertical component reduced by the fact that
the radiation flux exactly balances gravity in the equilibrium.  The
last three equations are the three components of the radiation flux equation.
These equations may be further simplified by the assumption (verifiable
{\it post hoc}) that $\omega \ll \kappa\rho c$; that is, the wave frequency
is very small compared to the mean time between photon scattering events.

   With the further approximation that $p_r \ll \rho c^2$, equations (5) --
(12) may be manipulated to yield a dispersion relation.  This relation is most
conveniently displayed in terms of dimensionless quantities, so that
$\omega = \omega_* \sqrt{g/h}$ and $\vec k = \vec k_{*}/h$.  To satisfy
the WKB approximation, $k_{*z} \gg 1$ and $k_{*r}\gg h/r$.
The dispersion relation is then seen to depend on six dimensionless
combinations of parameters:
\begin{eqnarray}
P & = & {p_r \over \rho g h} \\
G & = & {\sqrt{gh} \over c} \\
\Omega_{*}^2 & = & h\Omega^2/g \\
\tau & = & \kappa\rho h\\
{\cal Q}_p & = & (\partial Q/\partial p_r) (h/g)^{1/2} \\
{\cal Q}_\rho & = & (\partial Q/\partial \rho) (h g^3)^{-1/2}
\end{eqnarray}
Note that here $g$ is $g=g_z(z)$, the {\it local} value of the vertical
gravity.  In any thin disk, the quantity $G \ll 1$, for it is of order the
free-fall speed from the top of the disk to the midplane in units of $c$.
The relative importance of rotational effects is given by
$\Omega_{*}^2$, which is simply $h/z$ for Keplerian rotation.  The parameter
$P$ is also a function of height above the midplane.  In an optically
thick disk, $P = {1\over 2}[1 + 1/\tau - (z/h)^2]/(z/h)$.  Thus, the two parameters
$P$ and $\Omega_{*}^2$ could be replaced by the single parameter
$z/h$.  The last two dimensionless parameters describe the
sensitivity of the dissipation to the radiation pressure and the density,
respectively.  If one could extrapolate the ``$\alpha$" prescription to
local fluctuations, one might expect that ${\cal Q}_p \simeq \alpha$.

\section{The Dispersion Relation}

\subsection{General considerations}

In terms of these dimensionless quantities, the dispersion relation is:
\begin{eqnarray*}
\lefteqn{3\omega_{*}^5 
+ \left[i k_{*}^2 /(G\tau) - 3 k_{*z}G/\tau -i {\cal Q}_p \right]\omega_{*}^4}\\
& - & \left(4k_{*}^2 P + 3\Omega_{*}^2 + ik_{*z}^3/\tau^2
- ik_{*z} G {\cal Q}_p /\tau\right)\omega_{*}^3 \\
& + & \left[4k_{*z}^3PG/\tau - ik_{*}^2 \Omega_{*}^2/(G\tau)
+ i{\cal Q}_p (-ik_{*z} + \Omega_{*}^2) 
- i k_{*}^2 {\cal Q}_\rho \right]\omega_{*}^2 \\
& + & \left[4k_{*z}^2P\Omega_{*}^2 - 3k_{*r}^2 + k_{*z}^2 G(ik_{*z}{\cal Q}_\rho
- {\cal Q}_p)/\tau \right]\omega_* \\
& + & k_{*z}k_{*r}^2\Omega_{*}^2/(G\tau) + k_{*z}\Omega_{*}^2 (ik_{*z} {\cal Q}_\rho
- {\cal Q}_p) = 0 .\\
\end{eqnarray*}

    In the limit of $\tau \rightarrow \infty$ and ${\cal Q}_p = {\cal Q}_\rho
= 0$, this dispersion relation simplifies to
\begin{equation}
\omega_{*}^5 - \left[(4/3)k_{*}^2 P + \Omega_{*}^2\right]\omega_{*}^3 + 
\left[ (4/3) k_{*z}^2 P\Omega_{*}^2 - k_{*r}^2\right] \omega_* = 0.
\end{equation}
One root is clearly $\omega_* = 0$.  The other four are given by:
\begin{equation}
\omega_{*}^2 = \cases{ (4/3)k_{*}^2 P \cr
              \left[2 k_{*z}^2 \Omega_{*}^2 - (3/2)k_{*r}^2/P \right]/ k_{*}^2 \cr}
\end{equation}
The first pair of roots are the familiar radiation-supported
sound waves.  The second pair describe buoyancy behavior (cf. Balbus 1999).
If $k_{*z}^2 \Omega_{*}^2 \geq (3/4) k_{*r}^2/P$, there are two neutrally stable
gravity (epicyclic) waves; on the other hand,
if $k_{*z}^2 \Omega_{*}^2 < (3/4) k_{*r}^2/P$,
one mode is damped, but the other grows exponentially with essentially
no oscillation.  It is this last mode that corresponds to convection.
Although rotation tends to have a stabilizing effect, when $\Omega_*>0$
one can always find a mode with $k_r/k_z$ large enough to
satisfy this criterion.  As a consequence, when radiation pressure 
dominates gas pressure in a very optically thick disk, the standard equilibrium
is always unstable to convection.  However, relatively strong rotational
effects (i.e., $\Omega_{*}^2 P \sim (h/z)^2$ relatively large, or location
near the midplane) do diminish the range of wavevector directions that is
unstable.

\subsection{Dependence on parameters}

    As already remarked, both $P$ and $\Omega_*$ are determined by
$z/h$, and both should be, except very near the midplane, $\sim 1$.
Because the growth rate of the fastest growing mode in the optically
thick limit is $\simeq P^{-1/2}$, and $P$ decreases upward, convective
instability should develop most rapidly near the top of the disk.  This
expectation is borne out when the full equations are solved: as predicted
by equation 20, when $k_r \gg k_z$, the growth rate is nearly independent
of $|\vec k|$ and is $\simeq 7 \times$ greater at $z/h = 0.9$ than at
$z/h = 0.1$.

    The optical depth $\tau$ in a radiation pressure-dominated Shakura-Sunyaev
disk is
\begin{equation}
\tau = {4/3 \over \alpha} \dot m^{-1} x^{3/2} 
\left({\kappa \over \kappa_T}\right)^{-1} {R_z R_T \over R_{R}^2} ,
\end{equation}
where $\alpha$ is the usual dimensionless stress parameter, $\dot m$ is
the accretion rate in Eddington units for unit efficiency, $x$ is the
radius in units of $GM/c^2$, $\kappa/\kappa_T$ is the opacity relative
to pure Thomson opacity, and $R_{R,T,z}$ are the relativistic corrections
to the dissipation rate, torque, and vertical gravity (Page \& Thorne 1974;
Abramowicz, Lanza \& Percival 1997; see also Agol \& Krolik 2000a).  Because
we expect $\alpha, \dot m < 1$, but $x$ must be $> 1$, these disks should
be optically thick.  However, in the inner part of the disk, $\tau$
might be as little as $\sim 10^2$.
 
     Fig.~1 shows how the growth rate and phase velocity of the unstable
convective modes depend on wavenumber and $k_z/k_r$ when there is no
perturbation to the dissipation rate and the optical depth is very large
($\tau = 10^6$).  To illustrate the dependence on optical depth, the same
curves are shown in Fig.~2 for a case with the smallest optical depth one might
expect, $\tau = 10^2$.  In both cases, the range of meaningful wavenumbers
(and hence the range of wavenumbers displayed in the figures) is limited by two criteria.  On the one hand, the WKB approximation demands
that $k_z h \gg 1$; on the other, the diffusion approximation is only
consistent with $kh \ll \tau $.  We present results extending out to
$kh \sim \tau$; at the highest wavenumbers shown, the diffusion approximation
has only marginal validity.

    For fixed optical depth, the growth rate depends on $k_z/k_r$;
when $k_z/k_r \simgt 1.2$, the unstable convective mode disappears.
For smaller $k_z/k_r$ ($ < 0.1$), the growth rate is independent of
$k_z/k_r$.  Decreasing the optical depth reduces the extent of the ``flat'' portion of the growth rate curve for a given $k_z/k_r$.  However, the
largest value of the dimensionless growth rate, $(\omega_{*i})_{max}$,
in the wavenumber range of physical interest is almost independent of
$\tau$ over this range.

Figs. 1 and 2 also show the phase velocity of the unstable convective mode, normalized to the free fall speed in the disk.  For large optical depths,
the real frequency of convective modes is so close to zero that they can 
be considered as non-oscillatory; decreasing the optical depth towards the
smallest value ($\tau=10^2$) expected for radiation pressure dominated disks
leads to an increase in the real frequency of the mode, endowing it with a non-negligible phase velocity and, as a consequence, it becomes oscillatory.

In the absence of perturbations to the dissipation rate, finite (although
large) optical depth primarily affects the radiation sound waves (cf.
Agol \& Krolik 1998).  Not surprisingly, as the opacity falls, short
wavelength sound waves become damped.  Less intuitively, convective
modes remain unstable
even when the diffusion rate ($\sim k_*^2 c/(h\tau)$)
is larger than the characteristic frequency $\sqrt{g/h}$, i.e. when
$k_* > \sqrt{G\tau}$.  In this regime, epicyclic motions cause convective
modes to acquire an oscillatory character.
Convection remains unstable even in the face of strong
radiative diffusion because the equilibrium gas density is constant.  A
parcel that begins to fall because its specific entropy is too small
is squeezed as it encounters higher pressure.  Its radiation pressure
rises, but because diffusion is so effective, only up to the ambient
level.  Meanwhile, however, the increased gas density means that the
parcel's density exceeds that of its neighbors, and it continues to fall.

    Such impunity to the effects of radiation diffusion does not persist
all the way to zero opacity, however.  When $k_* > \tau$, the diffusion
approximation is no longer a valid description of radiation transfer.  For
wavelengths that short (or opacity that small), the radiation streams
freely, entirely independent of gas motions.  In that limit, if there
is nothing to alter the distribution of dissipation, there is no perturbation
to the flux as a function of position, and therefore, no perturbation to
the force.  The result, of course, is that convection disappears.

    Again applying the Shakura-Sunyaev disk solution, we expect $G$ to
be small, for it is given by
\begin{equation}
G = {3 \over 2} \dot m \left({\kappa \over \kappa_T}\right)
 \left({z \over h}\right)^{1/2} x^{-3/2} {R_R \over R_{z}^{1/2}} .
\end{equation}
$G$ becomes irrelevant over most of the wavenumber interval of interest when $\tau \rightarrow \infty$, as
every factor in which it appears in the dispersion relation
is divided by $\tau$; the value of $G$ can still have some influence 
even at large $\tau$ only for very large wavenumbers, $k_* \simgt 0.01\tau$,  
since some of those same terms of the dispersion relation actually are 
$\propto k_*^2/\tau$ or even $\propto k_*^3/\tau$.  However, even when
$\tau$ is as large as $10^5$, some dependence on $G$ begins to appear
across the range of interesting wavenumbers.
When the optical depth is no larger than this, diminishing $G$ tends to
enhance instability, particularly at short wavelengths.  When $\tau \sim
10^5$, the growth rate of the very short wavelength convective modes
increases by factors of several when $G$ falls from $10^{-2}$ to $10^{-4}$.
When $\tau \simlt 10^4$, and $G \sim 10^{-4}$, 
a second mode, in addition to the convective mode, becomes unstable 
at short wavelengths.  This mode is weakly
oscillatory, and its very low phase velocity ($\sim 2.5\times10^{-2}$ --
$1.6\times 10^{-3}$ the free fall velocity) behaves similarly to that of 
damped radiation sound waves at the same wavelengths.
Analysis of the corresponding amplitudes of the various physical quantitities'
perturbations shows that this mode has magnitude and especially phase
relationships between perturbations in pressure, 
density and velocity components that are intermediate between those 
of convective modes and those characteristic of radiation sound waves.

   It is important to note, however, that if the canonical disk
solution applies, the product
\begin{equation}
G\tau = 2 \left({z \over h}\right)^{1/2} \alpha^{-1} {R_T R_z^{1/2} \over R_R} .
\end{equation}
In other words, the product $G\tau$ should be very nearly independent of
radius (if $\alpha$ is).  Even where the relativistic corrections
are substantial, $G\tau$ hardly changes because $R_R$ and $R_T$ are
very nearly proportional,
and $R_z$ has a total range of at most a factor of two (Krolik 1999).
Thus, the criterion for which scaled wavelengths should be substantially
affected by photon diffusion (i.e., $k_* > \sqrt{G\tau}$) is almost
independent of radius, but does depend somewhat on $z/h$.

    Given our estimate of $G$ (equation 22), we can also verify our
expectation that
$\omega/c\kappa\rho \ll 1$.  Rewritten in terms of our dimensionless
parameters, this ratio is $\omega_* G/\tau$.  As we have just estimated,
$G \ll 1$ and $\tau \gg 1$ in radiation pressure-supported $\alpha$-disks.
Consequently, unless $\omega_*$ is extremely large, $\omega/c\kappa\rho \ll 1$
is a very safe assumption.

   As already pointed out, ${\cal Q}_p$ is much like a local version of the
vertically-averaged quantity $\alpha$, the ratio of the $r$--$\phi$
stress to the total pressure.  On this basis we expect ${\cal Q}_p$ to be
somewhat less than unity. The (dimensional) dissipation perturbation
coefficient $Q_{\rho}\equiv (\partial Q/\partial \rho)$, on the other hand, is
most closely related to the ratio of the flux to the surface density.  In a
Shakura-Sunyaev disk, this quantity (reduced to a dimensionless number
according to equation 18), is
\begin{equation}
\left({\tilde{F} \over \tilde{\Sigma}}\right)\equiv {{\cal F}\over \Sigma}{1\over (hg^3)^{1/2}} 
= {1 \over 2}\left({3 \over 2}\right)^{3/2} \alpha \dot m^{3/2}
\left({\kappa \over \kappa_T}\right)^{3/2} x^{-3/2} {R_{R}^{5/2} \over
R_{z}^2 R_T} .
\end{equation}
Although we don't know for certain that ${\cal Q}_\rho \sim \tilde{F}/\tilde{\Sigma}$,
it is a likely estimator.  On this basis, we expect ${\cal Q}_\rho \ll 1$,
and should be particularly small at larger radius or in disks with relatively
low accretion rate.

Like $\tau$, varying ${\cal Q}_p$ or ${\cal Q}_\rho$ affects the radiation sound
waves much more than the convective modes.  The growth rate of the
near-zero real frequency convective mode is almost independent of ${\cal Q}_p$
and ${\cal Q}_\rho$ over a large range in these parameters, but the longer-wavelength radiation sound waves become unstable when these
parameters are positive.  For example, when $z/h = 0.5$, and $\tau \gg 1$, 
the growth rates for positive ${\cal Q}_p$ and null ${\cal Q}_{\rho}$
are $\simeq (0.1-0.16){\cal Q}_p$, and, similarly, those for
positive ${\cal Q}_{\rho}$ and null ${\cal Q}_p$ turn out to be
$\simeq (0.1-0.16){\cal Q}_{\rho}$.

Unlike the convective mode, for radiation sound waves,
growth persists to $k_r/k_z < 1$.   Although the maximum growth rate does 
not depend on the value of this ratio, the range of wavelengths
for which growth occurs shrinks and moves towards larger wavelengths
as the ratio $k_r/k_z$ decreases.

    Smaller optical depth can also counteract the instability created
by positive ${\cal Q}_p$ or ${\cal Q}_\rho$ (see also Agol \& Krolik 1998
for a discussion of radiation diffusion damping of MHD waves).
In the very large $\tau$ limit, the maximum growth rate for
radiation sound waves is almost independent of the value of the optical depth, 
but for $\tau\simlt 10^4$ it decreases with decreasing $\tau$.
The radiation sound waves are all
damped for $\tau < 10^3$ or so; at larger optical depth, the shorter
wavelength modes remain damped (because, of course, diffusion is
most effective acting on them), but growth appears for longer wavelengths,
with the range of growing wavelengths stretching as $\tau$ increases.
Fig.~3 shows the behaviour of the imaginary part of the frequency and 
the normalised phase velocity for radiation sound waves with ${\cal Q}_p = 1$
and ${\cal Q}_\rho = 0$ as functions of wavenumber for a variety of optical depths.

    Summarizing this discussion, we conclude that there are two sorts of
instabilities of importance in radiation pressure-supported disks:
convective instabilities, which grow on roughly the dynamical timescale,
should be very nearly ubiquitous unless the disk becomes almost optically
thin; and radiation sound waves, which can be driven unstable (at a
slower rate) if ${\cal Q}_p$ or ${\cal Q}_\rho$ is positive.

\subsection{Comparison with the Classical Convection Problem}

    In standard treatments of convection in rotating fluids (e.g., Chandrasekhar
1961), it is shown that convection begins when the ``Rayleigh number"
\begin{equation}
{\cal R} = {g \alpha_V \beta h^4 \over \kappa_{th} \nu}
\end{equation}
exceeds a critical value (generally $\gg 1$) which depends on the ``Taylor
number"
\begin{equation}
{\cal T} = 4{\Omega^2 h^4 \over \nu^2} .
\end{equation}
Writing $T$ for the temperature, we define $\alpha_V \equiv |\partial\ln\rho/\partial T|$, $\beta \equiv |dT/dz|$,
$\kappa_{th}$ is the thermal conductivity, and $\nu$ is the kinematic viscosity.

    This analysis applies partially, but not entirely, to the circumstances of
radiation-dominated accretion disks.  The fact that, rather than entering
the fluid solely at the bottom, heat is actually generated throughout the
fluid probably alters the quantitative criteria for convective instability but
should not change anything at a qualitative level.  On the other
hand, certain contrasts are of greater importance.

    One that requires mention is that the pressure in these circumstances,
unlike in classical fluids, depends only on the radiation,
and is therefore independent of density.  Consequently, thermal diffusion
can alter the pressure with no change in density.  In the classical analog, thermal diffusion can quench convective instability by adding heat to falling
low-entropy fluid elements, thereby forcing them to expand until
their density no longer exceeds the surrounding density.  This occurs
when $\kappa_{th}$ is large enough that ${\cal R}$ falls below the
critical value.  By contrast, in the radiation-dominated case,
the density remains unchanged in the face of thermal diffusion.  This is
why convection persists even when $k_* > \sqrt{G\tau}$; i.e., the
radiation analog of the Rayleigh number does not accurately predict
the effect of thermal diffusion on the instability.

     Perhaps the most important contrast with classical fluids
is that there is no easy definition of the kinematic viscosity.
Bulk orbital shear leads to angular momentum transfer via magnetic forces
(Balbus \& Hawley 1998); these are, of course, entirely absent in our
treatment.  However, stresses arising from other shears may be
quite different, whether they are caused by magnetic forces or other
mechanisms (for example, photon diffusion: Agol \& Krolik 1998).  For this
reason, it is unclear whether the conventional equation of the $r-\phi$
shear stress with a nominal viscosity (as in the Shakura-Sunyaev $\alpha$
formalism) is appropriate here.  In any event, none of these possible
viscosity mechanisms is present in our equations; not even radiation viscosity
is considered, for no radiation shear stress exists in the pure diffusion
approximation.  Thus, in a formal sense, neither ${\cal R}$ nor
${\cal T}$ exists in our treatment of this problem.

    Nonetheless, having expressed these {\it caveats}, it is still of some
interest to estimate what ${\cal T}$ and ${\cal R}$ may be in radiation-dominated accretion disks in order to make some
contact with the classical theory.  To do so we make the suspect identification
discussed in the preceding paragraph: we suppose that the shear viscosity
in all directions is given by the Shakura-Sunayev $\alpha$.  If so, 
${\cal T} \simeq 10 \alpha^{-2}$, independent of radius (or anything else).
Within the radiation-dominated regime, the Rayleigh number is similarly
constant: ${\cal R} \sim O(10) (\kappa/\kappa_T)^2$.  Interestingly, the
critical ${\cal R}$ for $10 < {\cal T} < 10^5$ is $\sim 10^3$ -- $10^4$
(Chandrasekhar 1961).  Thus, if the most appropriate analog to ordinary
fluid viscosity is correctly estimated by this means, the strong
convective instability we find might be quenched; however, as we have stressed,
it is by no means clear that this analogy applies.

\section{Implications}

    We have shown that if an accretion disk is placed in the Shakura-Sunyaev
radiation-dominated equilibrium, the growth rate of the convective instability
is hardly affected by large changes in parameters such as the opacity or
the way in which the local dissipation might be perturbed.  This behavior
is, perhaps, not surprising, when one considers that the convective instability
is essentially dynamical, whereas the opacity and dissipation rates pertain
to the (rather more slowly accomplished) thermal balance.

    Given that the Shakura-Sunyaev equilibrium is unstable to convection
for all reasonable parameters, we can hardly expect that it describes
the actual state of radiation pressure-dominated disks.  One's first
guess might be that what happens instead is that entropy is redistributed
so as to make the disks at most marginally unstable to convection
(Bisnovatyi-Kogan \& Blinnikov 1977).  If the entropy distribution is at most
weakly unstable to convection, then departures from hydrostatic equilbrium
should be small, and the equilibrium could be determined by applying these
two conditions (i.e. isentropy and hydrostatic balance).  However, marginal
stability to convection also suggests that the amount of heat carried by
convection is small, so that radiative diffusion dominates energy exchange.
Unfortunately, this condition is inconsistent with the first two.  Any
two of these three conditions suffice to determine the equilibrium; the
solution is over-determined if all three are applied.  The
result must be, then, that in real disks none of them is satisfied exactly,
but all three are approximately correct.  The quantitative character of
this balancing act can be determined only by detailed calculations that
carry convection into the nonlinear regime (e.g. Agol \& Krolik 2000b).

    Deviation from constant specific entropy can certainly be expected
near the disk surface.  There, radiation diffusion should be able
to consistently dominate convection as a heat-loss mechanism because the
diffusion time is much shorter than in the body of the disk.  Consequently,
the specific entropy distribution near the surface must be clearly stable, i.e.
rise with increasing altitude.

    The sense of heat redistribution by convection is to raise the specific
entropy near the top and diminish it below.  For the same reason that
radiative diffusion must dominate near the top, this redistribution can
have the side effect of changing the overall rate at which heat is able
to leave the disk.  The mean specific entropy in equilibrium is therefore not
necessarily the same as the value predicted by the original Shakura-Sunyaev
equilibrium.

     These remarks do not bear on whether the thermal and viscous
instabilities continue to plague radiation-dominated disks.  It is possible
that convection might be able to remove extra heat generated in the
course of the thermal instability, and might consequently be able to
quench it.  Whether that is so can only be determined by detailed
calculation of heat transport by convection in the nonlinear regime
(Agol \& Krolik 2000b).
Whether the viscous instability acts under these conditions depends
on the character of dynamical coupling (see, e.g., Agol \& Krolik
1998) between the radiation and the MHD turbulence that most likely
accounts for angular momentum transport in disks (Balbus \& Hawley 1998).

     We have also shown that radiation sound waves may be driven unstable
if the emission of radiation is proportional to either the gas density or
the local radiation pressure (the latter possibility is suggested by a
straight-forward extrapolation of the ``$\alpha$-model).  Although
phenomenological guesses of this variety are plausible, we do not yet know
enough about the physical mechanisms of dissipation in accretion disks to
say whether this is what should actually happen.  However, if it does occur,
growth of radiation sound waves (and, presumably, magnetosonic waves) could
provide an additional source of turbulence in disks, supplementing
magneto-rotational instability and convection.  Indeed, if
dissipation leads to renewed stirring, there might be interesting twists
to the character of the MHD turbulent cascade.

\clearpage
\figcaption{Dispersion curves of unstable convection modes for very large
optical depth ($\tau = 10^6$). Non-dimensional growth rate,
$\omega_{*i}\equiv \omega_i/\sqrt{g/h}$, 
and logarithm of absolute value of non-dimensional phase 
velocity, $u_{ph}\equiv v_{ph}/\sqrt{gh} = \omega_{*r}/k_*$, are shown as functions of the non-dimensional wavenumber $k_*=kh$. In both plots the curves
are labelled with the corresponding value of the ratio of vertical to radial wavenumber $k_z/k_r$;
curves for $k_z/k_r <0.1$ are not shown, since, in the physically meaningful
wavenumber range, the growth rates are nearly identical to those found for $k_z/k_r=0.1$.}
\figcaption{Same as in Fig.~1, but for the smallest value of the optical depth 
expectable in the disk regime under examination, that is 
$\tau=10^2$. The range of non-dimensional wavenumbers shown is much smaller than
for Fig.~1 because of the restriction imposed by using the diffusion approximation.}
\figcaption{Dispersion curves for radiation sound waves in a case including 
perturbation of dissipation rate. The curves shown all refer to  
$k_z/k_r =0.7$ as a reference value; see the text for discussion of the
dependence of the dispersion curves on the value of the ratio $k_z/k_r$.
Different curves are labelled with the corresponding value of the optical depth
$\tau$. The wavenumber interval over which the mode is unstable depends on
$\tau$, growing in extent towards larger wavenumbers with increasing $\tau$.}

\begin{thebibliography}{}

\bibitem[Agol \& Krolik 1998]{ak98} Agol, E. \& Krolik, J.H. 1998, Ap. J.
507, 304

\bibitem[Agol \& Krolik 2000a]{ak00a} Agol, E. \& Krolik, J.H. 2000, Ap. J.
528, 161

\bibitem[Agol \& Krolik 2000b]{ak00b} Agol, E. \& Krolik, J.H. 2000, in
preparation

\bibitem[Balbus 1999]{b99} Balbus, S.A. 1999, astro-ph/9906315

\bibitem[Balbus \& Hawley 1998]{bh98} Balbus, S.A. \& Hawley, J.F. 1998,
Rev. Mod. Phys. 70,1

\bibitem[Bisnovatyi-Kogan \& Blinnikov 1977]{bkb77} Bisnovatyi-Kogan, G.
\& Blinnikov 1977, A. \& A. 59, 111.

\bibitem[Chandrasekhar 1961]{c61} Chandrasekhar, S. 1961, Hydrodynamic and Hydromagnetic Stability (Oxford UK: Oxford University Press)

\bibitem[Coroniti 1981]{c81} Coroniti, F. 1981,  Ap. J. 244,5887

\bibitem[Gammie 1998]{g98} Gammie. C. 1998, M.N.R.A.S. 297, 929

\bibitem[Krolik 1998]{k98} Krolik, J.H. 1998, Ap. J. Letts. 498, L13

\bibitem[Krolik 1999]{k99} Krolik, J.H. 1999, Active Galactic Nuclei: From 
the Central Black Hole to the Galactic Environment (Princeton: Princeton
University Press)

\bibitem[Liang 1977]{l77} Liang, E.P.T. 1977, Ap. J. Letts. 218, 243

\bibitem[Lightman \& Eardley 1974]{le74} Lightman, A.P. \& Eardley, D.M.
1974, Ap. J. Letts. 187, L1

\bibitem[Lominadze \& Chagelishvili]{cl84} Lominadze, Dzh. G. \& Chagelishvili,
G.D. 1984, Sov. A.J. 28, 168

\bibitem[Shakura \& Sunyaev 1973]{ss73} Shakura, N.I. \& Sunyaev, R.A.
1973, A. \& A. 24, 337

\bibitem[Shakura \& Sunyaev 1976]{ss76} Shakura, N.I. \& Sunyaev, R.A.
1976, M.N.R.A.S. 175, 613

\bibitem[Shakura, Sunyaev \& Zilitinkevich 1978]{ssz78} Shakura, N.I.,
Sunyaev, R.A. \& Zilitinkevich, S.S. 1978, A. \& A. 62, 179

\bibitem[Shapiro, Lightman \& Eardley 1976]{sle76} Shapiro, S.L.,
Lightman, A.P. \& Eardley, D.M. 1976, Ap. J. 204, 187

\bibitem[Svensson \& Zdziarski]{sz94} Svensson, R. \& Zdziarski, A.
1994, Ap. J. 436, 599

\bibitem[Szuszkiewicz \& Miller 1997]{sm97} Szuszkiewicz, E. \& Miller, J.I.
1998, M.N.R.A.S. 298, 888

\end{thebibliography}
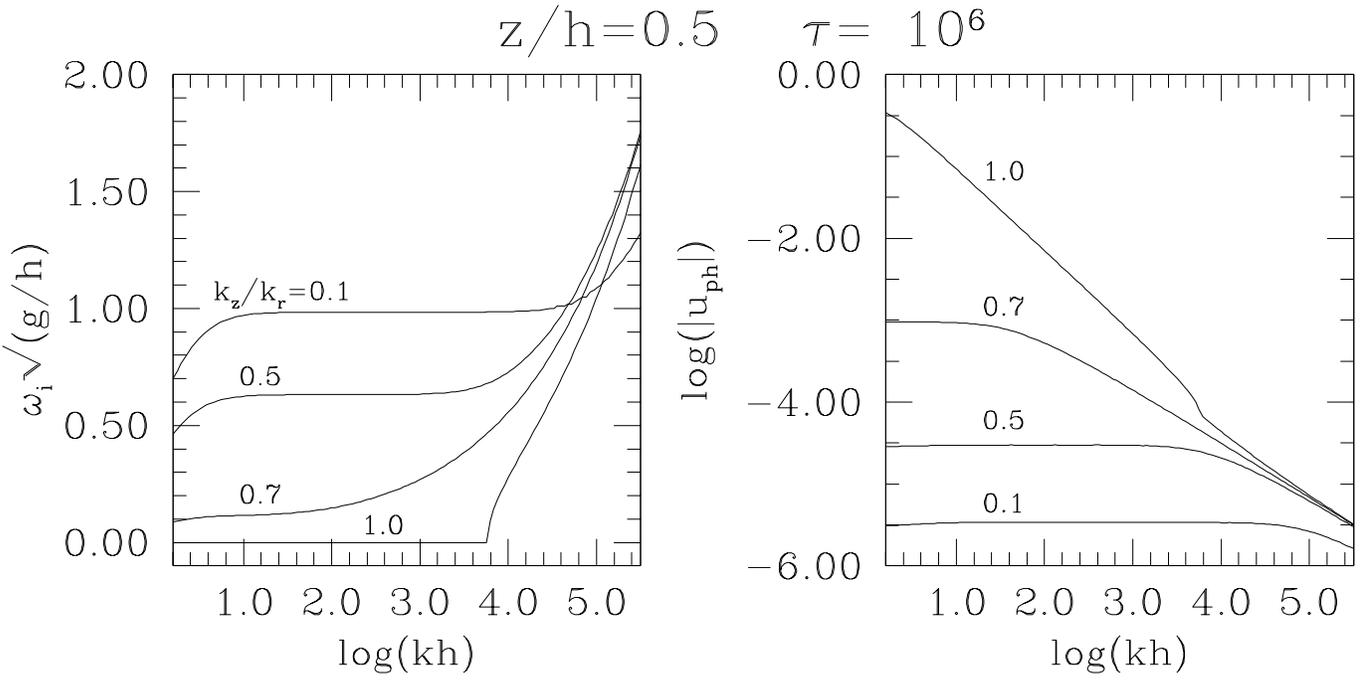
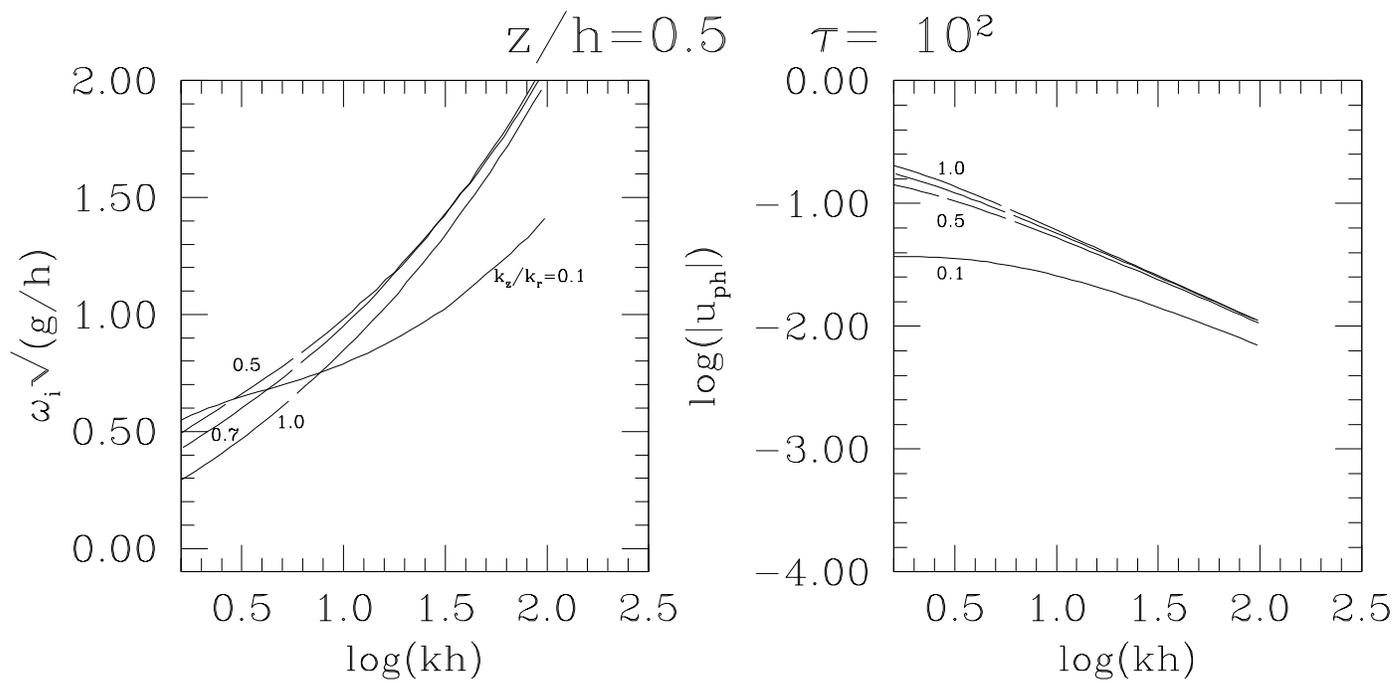
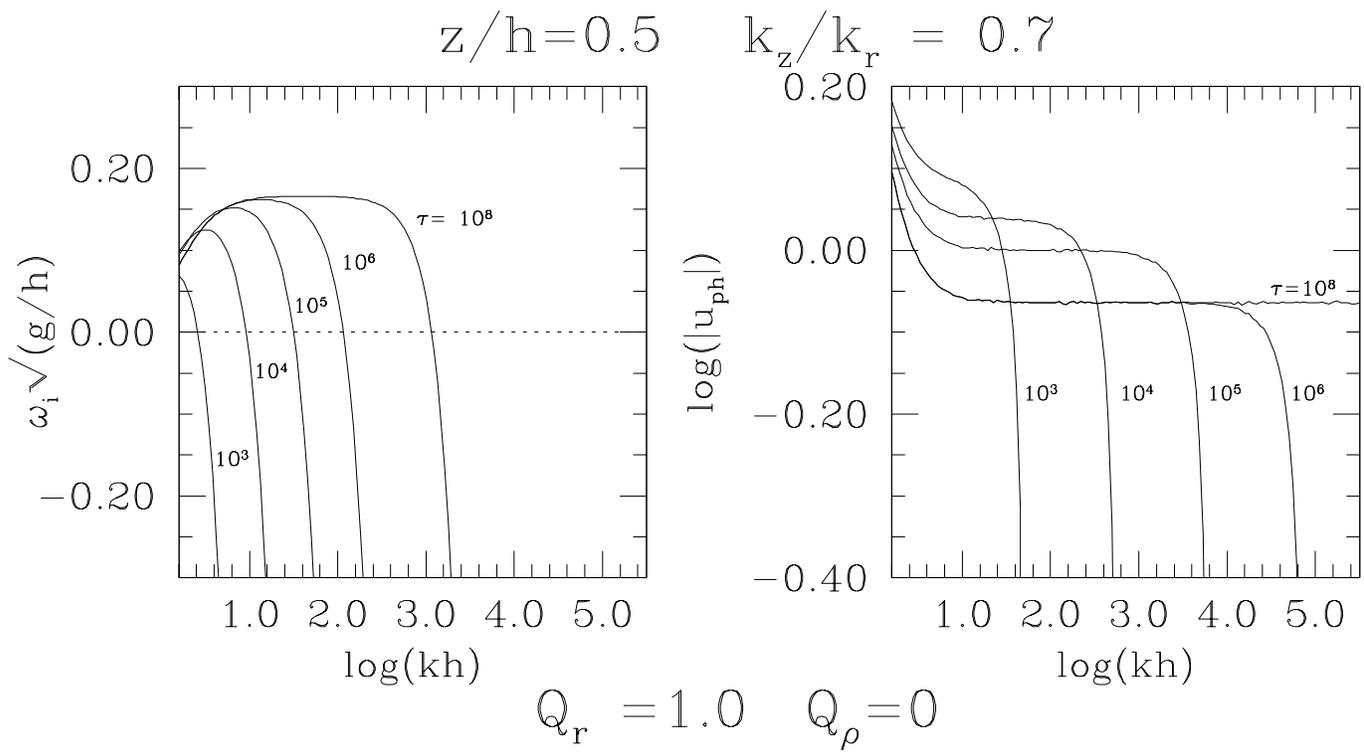
\end{document}